\documentclass[twocolumn,prl,aps,showpacs]{revtex4}
\draft
\begin{document}
\title{A robust Bayesian formulation of the optimal phase measurement problem}
\author{K. R. W. Jones\cite{**}}
\address{Sensors  Division, Alphaxon Research LLC,\\ 
Suite 11, Level 14 Lumley House, Sydney, Australia.}
\begin{abstract}
Optical phase measurement is a simple example of a quantum--limited measurement problem with important applications in metrology such as gravitational wave detection. The formulation of optimal strategies for such measurements is an important test--bed for the development of robust statistical methods for instrument evaluation. However, the class of possible distributions exhibits extreme pathologies not commonly encountered in conventional statistical analysis. To overcome these difficulties we reformulate the basic variational problem of optimal phase measurement within a Bayesian paradigm and employ the Shannon information as a robust figure of merit. Single--mode performance bounds are discussed, and we invoke a general theorem that reduces the problem of finding the multi--mode performance bounds  to the bounding of a single integral, without need of the central limit theorem.
\end{abstract}
\narrowtext
\maketitle

%\bigskip
Quantum limits to optical phase--shift measurement are
important in diverse areas; from the design of gravity 
wave detectors, to telecommunication, and optical 
fibre sensing. For the measured datum $\phi\in[-\pi,+\pi]$,
and an unknown ``true phase'' $\Phi$, the problem is
to achieve an {\em optimal statistical design\/}
for the phase detection curve $p(\phi|\Phi)$ 
which relates them (under a cost constraint
such as fixed average photon number, $N$).

The standard interferometric performance limits 
are well--known\cite{caves}: the shot--noise limit, 
$\Delta\phi<1/\sqrt{N}$, for coherent state inputs; 
and $\Delta\phi<1/N$, for optimized squeezed state 
inputs\cite{sqz}. However, under the stimulus provided 
by Shapiro, Shepard and Wong's\cite{ssw} suggestion
of a possible $O(1/N^{2})$ scheme, it has become
important to find a robust measure of optimality
that copes with statistical 
pathologies\cite{sdh,hall1,jones1,clb,bb}.

To paraphrase the overall problem, we may classify 
three basic tasks: 1) determine how to describe 
phase measurements; 2) determine how to prepare 
particular states---and implement the desired
measurements; and 3) determine, in company 
with the above, the best scheme under some 
chosen optimality criterion.

This letter concerns the last item, and so 
we pick a general theoretical setting due
to Shapiro and Shepard\cite{SS} that best
illustrates the difficulties.

Measurement is here described using the theory of
{\em probability operator measures\/}\cite{qet}. 
The {\em classical\/} data, say $\phi$, and the 
measured {\em quantum\/} state, say 
$\hat{\rho}(\Phi)$, are then related by the 
conditional probability rule
\begin{equation}
\label{pom}
p(\phi|\hat{\rho}(\Phi))\,d\phi = 
{\rm tr}[\hat{\rho}(\Phi)\hat{\Pi}(\phi)]\,d\phi,
\end{equation}
where $\hat{\Pi}(\phi)$ is a family of positive
hermitian operators which respects the closure 
constraint $\sum_{\phi} \hat{\Pi}(\phi)=\hat{1}$, 
so that $\sum_{\phi} p(\phi|\hat{\rho})=1$. The 
$\hat{\Pi}(\phi)$ need not be projectors (nor 
orthonormal, if they were).

For optical phase, Shapiro and Shepard\cite{SS} imagine
a scheme where an ingoing probe state is phase--shifted
by $e^{i\Phi\hat{N}}$ and then subjected to an idealized
Susskind--Glogower measurement\cite{SG}. For the rule
(\ref{pom}) they obtain
\begin{equation}
\label{ss}
p(\phi|\Phi) 
= \frac{1}{2\pi}\left|
\sum_{n=0}^{\infty} \psi_{n}
e^{i n(\phi-\Phi)}\right|^{2},
\end{equation}
where the $\psi_{n}$ are number--ket coefficients of 
the probe beam (to be optimized). Although there is
no known way to implement the SG--measurements, they
are thought to be optimal\cite{hall2} (and the derived 
statistics agree with the Pegg--Barnett hermitian phase 
operator approach\cite{PB}).

Interestingly, given some $p(\phi|\Phi)$, (\ref{ss})
can be ``inverted'' to find a corresponding minimum
average energy state\cite{SS}. Any kind of 
statistical behaviour is possible and one must
select a criterion that excludes pathologies. For instance, the 
SSW--state\cite{ssw} ($\psi_{n}=\sqrt{6}/\pi(1+n)$, 
for $n<M(N)$, or zero, with $M(N)$ chosen so $N$ 
is the mean photon number) is strictly optimal by 
reciprocal peak likelihood, but has been shown to 
be sub--optimal using other criteria for both 
single--mode\cite{sdh,hall1,jones1} and 
multi--mode\cite{clb} detection strategies.

One of the characteristic problems encountered in
such studies is to adequately evaluate the utility
of a sharp central peak sitting upon a broad tail.
It is this kind of pathology that the SSW--state
possesses. Measures such as peak likelihood, and 
rms--phase error bias one or other of these 
elements to a greater or lesser degree. As Hall
has argued\cite{hall1}, there are at least two
good candidates, the use of confidence
intervals, or the Shannon information (Fisher 
information\cite{clb} is another possibility 
for the analysis of multi--mode schemes).  
In a recent paper, Bialynicki--Birula et al.\cite{bb} 
reported a numerical optimization of the 
single--mode problem for {\em five\/} 
different criteria. They found that the 
Shannon information occupied the 
``middle ground'' among these. It seems not to 
place undue emphasis upon either ``peaks''
or ``tails'', which is desirable to fix 
a robust variational 
problem that rejects false solutions. 
Indeed Shannon information can exclude
even the most extreme pathologies,
such as a {\em singular peak\/} 
atop a {\em broad tail\/}\cite{jones1}. 

With the goal of robustness in mind we 
reformulate the general multi--mode
optimization problem in information 
theoretic terms. A secondary
purpose will be to show how the Bayesian
methodology fits easily with entropic
measures of uncertainty, and quantum
mechanics\cite{jones2}.

The important feature of (\ref{pom}) is that 
we can fix upon a $\hat{\rho}$, and imagine a 
class of all possible $\hat{\Pi}(\phi)$, or 
vice--versa. Some choice returns a 
$p(\phi|\hat{\rho})$. This rule is
read as $p({\rm data}|{\rm state})$, and 
we see that a ``good'' instrument must 
closely correlate particular {\em data\/} 
(i.e. the observed readings) with some
particular {\em state\/} (i.e. that we 
wish to know). Ideally, we seek a delta 
function correlation, but in general it 
will be more fuzzy due to the effects 
of quantum and classical noise.

In the Bayesian viewpoint\cite{daniell} one looks 
upon this link as being reflexive, i.e. we seek 
to find $p({\rm state}|{\rm data})$. To do this 
one must introduce a {\em prior probability\/}, 
$p_{0}({\rm state})$, for the as yet unknown 
states. Then we use Bayes' rule of conditionals:
\begin{equation}
\label{bayes} 
p({\rm data},{\rm state})\equiv
p({\rm data}|{\rm state})p_{0}({\rm state}).
\end{equation}
to perform the ``statistical inversion''
\begin{equation}
\label{inversion} 
p({\rm state}|{\rm data}) =
\frac{p({\rm data}|{\rm state})p_{0}({\rm state})}
{\sum_{{\rm state}} p({\rm data}|{\rm state})p_{0}({\rm state})}.
\end{equation}
In general, two problems arise. We may not 
know what $p({\rm data}|{\rm state})$ is, 
or we may not have a good way to single 
out a prior distribution $p_{0}({\rm state})$\cite{central}.

In quantum theory, the situation is better
than one might first expect\cite{jones2}. 
Now, unlike in classical statistics, we can engineer
a {\em particular\/} $p({\rm data}|{\rm state})$. It 
is subject to control, and design (as evidenced by 
the optimal phase measurement problem). Secondly, 
the space of states is a {\em physical space\/} 
upon which {\em physical\/} symmetry principles
can be brought to bear to fix the Laplacian 
notion of {\em a priori complete ignorance\/}. 
For optical phase the answer is obvious. We 
choose $p_{0}(\phi)=1/2\pi$, that unique 
function invariant under phase changes 
$\phi\mapsto \phi+\delta\phi$ (an example 
of a general principle advocated 
by Jaynes\cite{jaynes}).

Now it remains to quantify optimality. In 
general, we must place a figure of merit 
upon $p({\rm state},{\rm data})$, the
joint correlation between states and
data. Significantly, it is {\em not\/} 
merely $p({\rm data}|{\rm state})$ 
that matters. For instance, one can
imagine an instrument that was very 
accurate for some states, and poor 
for others. Optimal measurement is
thus a notion defined relative to 
{\em those situations we expect 
to encounter in practice\/}.

In the optimal design problem we must 
look, therefore, for a figure of merit
defined upon $p({\rm state},{\rm data})$, 
with $p_{0}({\rm state})$ chosen to
reflect our design intentions.

The standard measure of covariance, based
upon an analysis of variance, is the
obvious choice. However, to ensure a
robust solution we will employ the
{\em mutual information\/}\cite{shannon}
\begin{equation}
\label{mut}
\langle\Delta{\cal I}\rangle =
\sum_{\rm state}\sum_{\rm data}
p({\rm data},{\rm state})
\log_{2}\left[\frac{p({\rm data},{\rm state})}
{p({\rm data})p({\rm state})}\right],
\end{equation}
of communication theory. This quantity 
is non--negative, and zero if, and only 
if, the distributions are statistically 
independent (an uninformative 
measurement)\cite{jones2}.

Further, one has an obvious communication
theoretic analogy. The above measure is
the average number of {\em bits\/} that 
could be sent if we encoded messages in
a set of physical states that are sent 
with probability $p_{0}({\rm state})$
(in practice a relative frequuency).
Here it measures the information 
gained from data about the state, 
{\em for an instrument whose performance is 
assessed on an imaginary ensemble of states 
distributed according to\/} $p_{0}({\rm state})$. 

Now we apply (\ref{mut}) to the optical
phase measurement problem. Going back to 
(\ref{ss}) one may think of $\psi(\Phi)$ 
as the ``information carrier'', a phase
modulated signal, and set $p_{0}(\Phi)=1/2\pi$, 
so that all phase--shifts are equally 
likely {\em a priori\/}. Then the
$\sum_{\rm state}$ becomes integration $d\Phi$, 
with $\Phi\in[-\pi,\pi]$, and similarly for 
the $\sum_{\rm data}$.

From (\ref{ss}) we obtain the multi--mode correlation
\begin{equation}
\label{multi}
p(\phi_{1},\ldots,\phi_{m}|\Phi) \equiv
\prod_{k=1}^{m}\frac{1}{2\pi}
  \left|\sum_{n=0}^{\infty} \psi_{n} e^{in(\phi_{k}-\Phi)}\right|^{2}.
\end{equation}
Choosing the uniform prior $p_{0}(\Phi)=1/2\pi$, we 
apply Bayes' rule (\ref{bayes}) to obtain
\begin{eqnarray*}
p(\phi_{1},\ldots,\phi_{m})
& \equiv & \int_{-\pi}^{\pi}\,d\Phi
p(\phi_{1},\ldots,\phi_{m}|\Phi)p_{0}(\Phi)\nonumber \\
&    =   &
\int_{-\pi}^{\pi}\,\frac{d\Phi}{2\pi}
\prod_{k=1}^{m}\frac{1}{2\pi}
  \left|\sum_{n=0}^{\infty} \psi_{n} e^{in(\phi_{k}-\Phi)}\right|^{2}. 
\end{eqnarray*}
Then, using (\ref{inversion}), we have
\begin{equation}
p(\Phi|\phi_{1},\ldots,\phi_{m}) 
= \frac{p(\phi_{1},\ldots,\phi_{m}|\Phi)}
         {2\pi p(\phi_{1},\ldots,\phi_{m})}.
\end{equation}
Using (\ref{bayes}) once more, we substitute this
into (\ref{mut}), and rearrange to obtain
\begin{eqnarray}
\lefteqn{
\langle\Delta{\cal I}(N)\rangle
 = \int_{-\pi}^{\pi}\stackrel{(m)}{\ldots}\int_{-\pi}^{\pi}
   d\phi_{1}\ldots d\phi_{m}
   p(\phi_{1},\ldots,\phi_{m})   } \nonumber \\
& &
\int_{-\pi}^{\pi}\,d\Phi
p(\Phi|\phi_{1},\ldots,\phi_{m})
\log_{2} 
\left(2\pi
p(\Phi|\phi_{1},\ldots,\phi_{m})\right).
\label{ssm}
\end{eqnarray}
as the gain in {\em bits\/} for a multi--mode 
measurement on $m$ identical pulses yielding
the data $\phi_{1},\ldots,\phi_{m}$. In 
this problem one must optimize, cojointly,
the chosen $\psi_{n}$, and the number of
pulses $m$, subject to the total average 
photon number constraint $N = m \bar{n}$,
where $\bar{n} = \langle \hat{a}^{\dagger}\hat{a}\rangle_{\rm single mode\/}$,
is the average photon number per mode
(see Lane et al.\cite{clb}).

In the special case of a single--mode we set
$\theta =\phi-\Phi$, and introduce the new
function $f(\theta)= 2\pi p(\phi|\Phi)$, 
where $f(\theta)=f(-\theta)$, Then
(\ref{ssm}) assumes the simple form:
\begin{equation}
\label{ent}
\langle\Delta{\cal I}(N)\rangle
= \int_{-\pi}^{\pi}
f_{N}(\theta) \log_{2}f_{N}(\theta)\,d\theta/2\pi,
\end{equation}
where $N$ is the mean photon--number of the
single mode, and our interest lies in the 
regime $N\gg 1$, for parametric families
of probe states $\psi_{n}(N)$.

Elsewhere\cite{jones1}, we used (\ref{ent}) to
reconsider the three trial states of Shapiro, 
Shepard and Wong's paper\cite{ssw}. In their
naming scheme, we get:
\begin{mathletters}
\label{allc}
\begin{eqnarray}
\langle\Delta{\cal I}_{\rm SSW}(N)\rangle
& \le &
0.966 \label{bssw},\\
\langle\Delta{\cal I}_{\rm CS}(N)\rangle
& \sim &
1/2 \log_{2} N - 1.604 \label{bcs},\\
\langle\Delta{\cal I}_{\rm TS}(N)\rangle
& \sim &
\log_{2} N - 0.220\label{bts}.
\end{eqnarray}
\end{mathletters}
Whereas the coherent state (CS) and truncated 
phase state (TS) return unbounded information 
gain, we may expend {\em infinite\/} photon
energy and recover no more that {\em one 
bit\/} from the SSW--state.

This marked sub--optimal behaviour may be traced 
to the large--$N$ vanishing peak--area property 
of the SSW--state noted by Schleich et al.\cite{sdh}, 
or to its well--known extended tails\cite{ssw}. 
In yet another view, one can employ a simple 
scaling argument to explain this unusual 
finite--gain boundedness\cite{jones1}.

Although this example shows that (\ref{mut}) 
is a robust criterion, the optimization is 
now more difficult. In nonlinear problems 
of this kind, the simplest line of attack 
is to seek an upper bound. For the given
single--mode example, Hall\cite{hall1} 
has done this by adapting an entropic 
uncertainty relation\cite{birul}, to 
obtain the inequality:
\begin{equation}
\label{bound}
\langle\Delta{\cal I}(N)\rangle
 \le \log_{2}(N+1) + N\log_{2}(1 + 1/N)
\end{equation}
Comparing this with (\ref{bts}), Hall notes\cite{hall1}
that the truncated--phase (discrete--phase) states are
within $1.220$ bits of the theoretical optimum. To interpret
the physical meaning of such {\em pure numbers\/} we 
consider a typical asymptotic gain of the form\cite{jones1}
$$\langle\Delta{\cal I}(N)\rangle
\sim \log_{2}N + \beta
= \log_{2} (2^{\beta}N),$$
where $\beta<\beta_{\rm op}= 1$ (from (\ref{bound})). 
Define $\Delta \beta =\beta_{\rm op}-\beta$, and it
becomes clear that $N = 2^{\Delta \beta} N_{\rm op}$,
is the energy--expenditure conversion factor at fixed
information gain. Since $2^{\Delta \beta} \approx 
2^{1.220}= 2.329$, (\ref{bts}) is {\em twice\/} as
expensive as the optimal strategy (and there are 
a number of candidates with similar single--mode 
performance). For all practical purposes this is 
not so bad at all (contrast (\ref{bcs}), having
geometric inferiority).

Analysis of the multi--mode case envisaged in\cite{ssw},
is far more challenging. One must then account for
the problem of optimally choosing the partition
$N=\bar{n}m$. Recently, Lane et al.\cite{clb} 
showed, via exhaustive Monte--Carlo simulations 
of a maximum likelihood data analysis scheme,
that the effective multi--mode error 
scaling law is $O(1/N^{0.85})$ for an 
optimized SSW--partition. This is less 
than the $O(1/N^{2})$ Shapiro et al. 
had hoped for (and still inferior 
to squeezed--state interferometry),
but it shows that such avenues 
must be closed. 

On these grounds, we advocate the maximization 
of (\ref{ssm}) as a robust variational problem. 
Previously, the Fisher information\cite{fish}
was used as the optimality criterion\cite{clb}
(since that is the key tool in the analysis 
of variance for maximum likelihood
methods\cite{cramer,braun}). However, recent
work in the information theoretic asymptotics 
of Bayes methods\cite{barron} has shown that 
the Fisher and Bayes methods are essentially
equivalent for uniform prior in the large 
$m$ regime. Of course, only there is the 
theorem of Fisher valid anyway\cite{fish,cramer,braun}.

Thus we expect the two variational problems
will be asymptotically equivalent. Further,
as we will see, the criterion (\ref{ssm}) 
suggests the existence of multi--mode 
bounds analogous to Hall's single--mode
bound given at Eq.~(\ref{bound}).

The rationale for preferring (\ref{ssm}) in
this aim is as follows. A key difficulty in 
the maximum likelihood analysis\cite{clb},
is the huge computational cost posed by 
the open--ended multi--mode data set 
$\{\phi_{1},\phi_{2},\ldots\}$.  The
optimal division of pulse energy is
{\em unknown\/} a priori.

Further, in the multi--mode problem we must 
allow for {\em any\/} possible statistical 
behaviour, for both the large $m$ limit, 
and the case where $m=0(1)$ (where the 
general expectation seems to be that 
the optimal result occurs for $m=1$). 
This is problematic because one then
needs corrections to the Fisher
result, arising from the higher
order asymptotics of the central 
limit theorem\cite{braun}. 

Statistical methods to locate the transitional 
regime to the asymptotic normality predicted 
by Fisher\cite{fish} have been developed by
Braunstein\cite{braun}. While this is very
useful to estimate the true performance of
a multi--mode scheme\cite{clb}, one must
employ Monte--Carlo simulations to verify 
the domain of applicability anyway.

The new approach we advocate is to recognize
that the multi--mode performance is limited 
by the ``best possible'' statistical event
(irrespective of how likely it is; i.e. we
do not care if it is rare).

Examining Eq.~(\ref{ssm}) we see that 
$\langle\Delta{\cal I}(N)\rangle$ is bounded 
above by the posterior distribution
$p(\Phi|\phi_{1},\ldots,\phi_{m})$
of greatest information (i.e. we replace 
$p(\phi_{1},\ldots,\phi_{m})$ in (\ref{ssm})
by a delta function centered on this datum). 
It is perhaps intuitively clear (see later) 
that this is generated by the (very unlikely) 
identical data string\cite{jones3}
$$\{\phi_{1},\phi_{2},\ldots,\phi_{m}\}
= \{\phi,\phi,\ldots,\phi\/\},$$
since this is the ``most peaked'' possible
product of $m$ single--mode functions. In
the case of a uniform prior we can leave
$\phi$ {\em arbitrary\/}, since the
information is then independent of 
$\phi$. Specifically, we choose
\begin{equation}
\label{product}
p_{N}(\Phi|\phi_{1},\phi_{2},\ldots,\phi_{m})
={\cal N}(m)^{-1} [ p_{\bar{n}}(\phi|\Phi) ]^{m},
\end{equation}
with $\phi$ arbitrary, where the normalization is
$${\cal N}(m) = \int_{-\pi}^{+\pi}
[ p_{\bar{n}}(\phi|\Phi) ]^{m}\,d\Phi/2\pi,$$
and $p_{\bar{n}}(\phi|\Phi)$ is the single--mode
detection function.

Recently, we proved a theorem\cite{jones4} (under 
very general conditions that go beyond the present
application) that the {\em average\/} multi--mode 
performance of a fixed single--mode function is 
limited by the ``best--case'' result of $m$ 
{\em identical data\/}. This theorem implies 
that
\begin{equation}
\langle\Delta{\cal I}(N)\rangle
\le 
\int_{-\pi}^{+\pi}
\frac{[p_{\bar{n}}(\phi|\Phi) ]^{m}}{{\cal N}(m)}
\log_{2}\left(
2\pi
\frac{[p_{\bar{n}}(\phi|\Phi) ]^{m}}{{\cal N}(m)}
\right)
\,d\Phi,
\end{equation}
for the case of a uniform prior. Significantly,
we {\em do not need\/} the central limit
theorem to show this (it follows from a
convexity argument for {\em any\/} convex 
optimality measure, i.e. not just 
information\cite{jones3,jones4}). 

Thus, to locate an absolute {\em multi--mode\/}
performance bound, {\em for all $m$\/} (both
large and small), we need only study this 
single integral. Although the true average
performance must include a statistical 
analysis of {\em all\/} the outcomes, 
and their likelihood, we see that 
the setting of upper bounds does 
not require this. 

This is most helpful if, by the analysis
of bounds, we can {\em show\/} that the
multi--mode scheme cannot realize any
useful performance increase. This is
the expected result after the work
of Lane et al.\cite{clb}. 

The multi--mode problem thus becomes clearer, 
and a resolution of the issue is perhaps within 
sight. One would like to extend (\ref{bound}) 
so as to limit the Shannon information realized 
by an arbitrary product function (\ref{product}), 
where $\bar{n}$ is subject to the usual
constraint $N=m\bar{n}$. Although it 
remains difficult, this problem is 
more tractable than the maximum 
likelihood analysis, and may 
well be amenable to a direct
analytical assault. 

While a solution is always preferable to a
bound, the ``bounding strategy'' appears
to be the fastest route to discover if
multi--mode schemes are worth it. This
route offers hope that we can avoid
the central limit theorem corrections
needed in maximum likelihood
analysis\cite{braun}.

In conclusion, the optimal phase measurement
problem provides a challenge to the standard
methods based upon analysis of variance. If, 
as in this case, {\em all\/} conceivable
statistical functions are candidates in 
principle\cite{SS} new robust methods 
seem essential.

This work was sponsored by the Australian Research Council and was largely completed some twenty years ago. However, at that time, Bayesian methods were not widely understood and certainly not accepted within the physics community. The variational problem posed herein remains unsolved to this day.

\end{document}